\documentclass[runningheads]{llncs}
\usepackage[T1]{fontenc}
\usepackage{graphicx}
\usepackage{amsmath}
\usepackage{amsfonts}
\usepackage{amssymb}
\usepackage[ruled,vlined]{algorithm2e}
\usepackage{multirow}
\usepackage{array}
\usepackage{booktabs}
\usepackage{caption} 
\usepackage{float}
\usepackage{subcaption}
\usepackage[misc]{ifsym}
\usepackage{orcidlink}
\usepackage{bm}
\begin{document}
\title{PhishSSL: Self-Supervised Contrastive Learning for Phishing Website Detection}
\titlerunning{PhishSSL: Self-Supervised Contrastive Learning for Phishing Detection}
% If the paper title is too long for the running head, you can set
% an abbreviated paper title here
%
\author{WENHAO LI\inst{1}\orcidlink{0009-0007-4342-6676},
SELVAKUMAR MANICKAM\inst{1}\textsuperscript{\Letter}\orcidlink{0000-0003-4378-1954},
YUNG-WEY CHONG\inst{2}\orcidlink{0000-0003-1750-7441},
SHANKAR KARUPPAYAH\inst{1}\orcidlink{0000-0003-4801-6370},
PRIYADARSI NANDA\inst{3}\orcidlink{0000-0002-5748-155X} and
BINYONG LI\inst{4}\orcidlink{0000-0003-3615-1129}}

\authorrunning{W. Li et al.}
% First names are abbreviated in the running head.
% If there are more than two authors, 'et al.' is used.
%

\institute{
Cybersecurity Research Centre, Universiti Sains Malaysia, Pulau Pinang, Malaysia \\
\email{wenhaoli@ieee.org, \{selva, kshankar\}@usm.my} \and
School of Computer Sciences, Universiti Sains Malaysia, Pulau Pinang, Malaysia \\
\email{chong@usm.my} \and
Faculty of Engineering and IT, University of Technology Sydney, Sydney, Australia \\
\email{priyadarsi.nanda@uts.edu.au} \and
Chengdu University of Information Technology, Chengdu, China \\
\email{lby@cuit.edu.cn}}
\maketitle              % typeset the header of the contribution

\begin{abstract}
Phishing websites remain a persistent cybersecurity threat by mimicking legitimate sites to steal sensitive user information. Existing machine learning–based detection methods often rely on supervised learning with labeled data, which not only incurs substantial annotation costs but also limits adaptability to novel attack patterns. To address these challenges, we propose PhishSSL, a self-supervised contrastive learning framework that eliminates the need for labeled phishing data during training. PhishSSL combines hybrid tabular augmentation with adaptive feature attention to produce semantically consistent views and emphasize discriminative attributes. We evaluate PhishSSL on three phishing datasets with distinct feature compositions. Across all datasets, PhishSSL consistently outperforms unsupervised and self-supervised baselines, while ablation studies confirm the contribution of each component. Moreover, PhishSSL maintains robust performance despite the diversity of feature sets, highlighting its strong generalization and transferability. These results demonstrate that PhishSSL offers a promising solution for phishing website detection, particularly effective against evolving threats in dynamic Web environments.

\keywords{Anti-Phishing \and Contrastive Learning \and Phishing Website Detection \and Social Engineering \and Self-Supervised Learning.}
\end{abstract}
\section{Introduction}
Phishing websites remain one of the most pervasive and effective vectors in cyberattacks, posing significant threats to users by mimicking legitimate websites to steal sensitive information such as credentials, banking details, and personal data \cite{10788671}. As phishing campaigns grow increasingly sophisticated—employing evasive techniques like domain spoofing, dynamic content generation, and semantic mimicry—developing robust and generalizable phishing detection methods has become a pressing challenge for cybersecurity researchers and practitioners \cite{10175532}.

Existing phishing detection methods largely rely on supervised machine learning (ML) and deep learning (DL) models. Although these approaches demonstrate strong performance in benchmark settings, their dependence on labeled data and previously observed attack patterns leads to high annotation costs and poor adaptability. As a result, they often struggle to generalize to unseen phishing tactics and exhibit limited adaptability to evolving attack strategies. In contrast, self-supervised learning (SSL) has recently demonstrated remarkable success in related security domains, such as intrusion detection and malware classification, by reducing reliance on labeled data and improving generalization to unseen patterns.

Therefore, we propose PhishSSL, a novel self-supervised contrastive learning framework that requires no labeled data during training. Instead, PhishSSL is trained entirely on unlabeled data and leverages mixed tabular augmentations to generate semantically consistent positive samples. The model learns feature-based embeddings through contrastive learning and incorporates a learnable feature attention mechanism to emphasize discriminative patterns. Our main contributions are summarized as follows:
\begin{itemize}
    \item We design and implement PhishSSL, a self-supervised contrastive learning framework for phishing detection that operates without labeled datasets for training, integrating tabular augmentations and attention-based embedding.
    \item We conduct comprehensive comparisons with several unsupervised and self-supervised baselines on three phishing datasets, demonstrating the superior performance of PhishSSL.
    \item We perform an ablation analysis to evaluate the impact of core components such as feature attention, dropout augmentation, and traditional tabular augmentation.
\end{itemize}

The remainder of the paper is organized as follows: Sec.~\ref{related} reviews related work, Sec.~\ref{method} presents the proposed PhishSSL method, Sec.~\ref{eval} describes the evaluation setup, Sec.~\ref{results} discusses the results, and Sec.~\ref{conclusion} concludes the paper.

\section{Related Work}\label{related}
\textbf{Phishing Website Detection.}  
Traditional phishing website detection methods predominantly leverage supervised ML and DL models trained on handcrafted features extracted from URLs, HTML content, and domain metadata \cite{10788671}. Common approaches include Decision Trees, Support Vector Machines (SVMs), Convolutional Neural Networks (CNNs), Recurrent Neural Networks (RNNs), and ensemble techniques, which achieve high accuracy on benchmark datasets by modeling known phishing heuristics and statistical patterns \cite{almujahid2024comparative,kavya2024staying}. More recently, graph-based models such as Graph Neural Networks (GNNs) have been introduced to capture structural and relational dependencies between Web elements \cite{hal04401167}, further enhancing detection capability. However, these models typically require large amounts of labeled data, which can be expensive and time-consuming to obtain, particularly as phishing tactics continue to evolve. Moreover, their reliance on static training data limits adaptability, as attackers frequently alter content and structure to bypass detection mechanisms \cite{10175532}.

\textbf{Self-Supervised Contrastive Learning.} 
SSL has emerged as a compelling paradigm for representation learning without the need for manual labeling \cite{10559458}. Contrastive learning, a subclass of SSL, trains models to pull semantically similar samples closer and push dissimilar ones apart in the embedding space \cite{pmlr-v119-chen20j}, which has achieved significant success in vision and natural language processing.

Self-supervised contrastive learning has been widely applied in cybersecurity tasks. In intrusion detection, it enables the learning of robust flow- or packet-level representations from raw traffic, improving generalization across network domains \cite{koukoulis2025self}. In container security, contrastive learning has been used to reduce false alarms and detect runtime attacks without manual labeling \cite{10.1145/3665795}. In smart contract analysis, it captures semantic relationships between contracts to enhance vulnerability detection \cite{10.1145/3597503.3639173}. For in-vehicle networks, it supports anomaly detection in the Internet of Vehicles with high robustness under varying attack conditions \cite{10945136}. In Internet of Things (IoT) malware classification, contrastive learning improves accuracy by learning discriminative representations from transformed malware samples using limited labels \cite{WANG2025110299}.

While ML/DL-based methods have advanced phishing detection and SSL has demonstrated effectiveness in other security domains, contrastive SSL remains largely unexplored for phishing websites. Addressing this gap offers a promising avenue to minimize reliance on labels while improving both efficacy and generalization in phishing website detection.

\section{Methodology}\label{method}

\subsection{Framework Overview}
The fundamental principle underlying our approach is that legitimate and phishing websites exhibit distinct patterns in their structural, content, and external characteristics, even without explicit labels. By leveraging contrastive learning, the framework learns to distinguish between semantically similar augmented views of the same website while separating dissimilar samples in the embedding space. This approach eliminates the dependency on manual labeling while capturing intrinsic website properties that generalize across different phishing strategies.

Given an input feature vector $x_i \in \mathbb{R}^{87}$ representing comprehensive website characteristics (illustrated in Sec.~\ref{sec:feature}), the objective is to learn an embedding function $f(\cdot): \mathbb{R}^{87} \rightarrow \mathbb{R}^d$ that minimizes distances between semantically equivalent augmented views while maximizing distances between dissimilar samples. Formally, the learning objective can be expressed as:

\begin{equation}
\min_{f} \mathbb{E}_{x,x^+,x^-}[\max(0, d(f(x), f(x^+)) - d(f(x), f(x^-)) + m)]
\end{equation}

where $x^+$ represents positive (augmented) views, $x^-$ represents negative samples, $d(\cdot, \cdot)$ denotes distance metric, and $m$ is the margin parameter.

\begin{figure}[!ht]
\centering
\setlength{\belowcaptionskip}{-1cm}
\includegraphics[width=\linewidth]{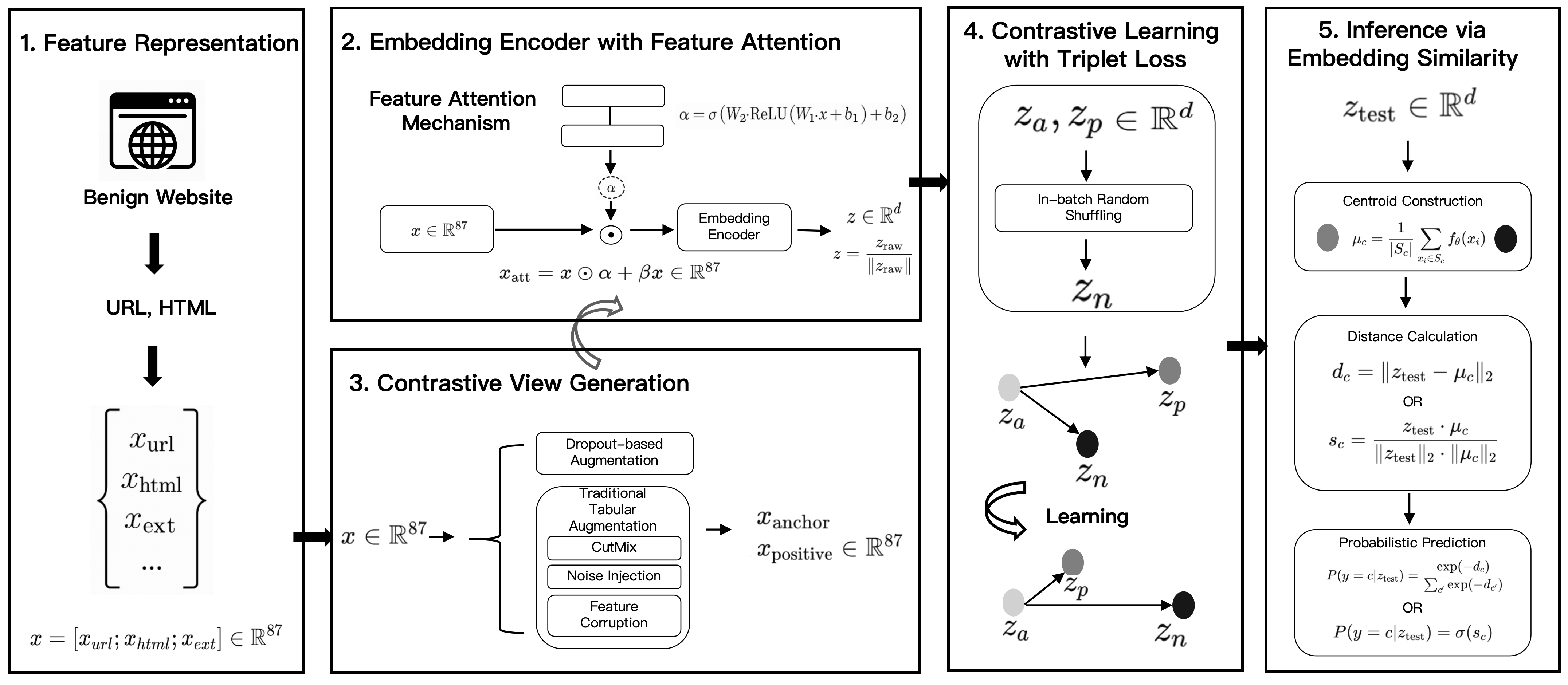}
\caption{Proposed PhishSSL Framework.}
\label{fig:phishssl}
\end{figure}

The Fig.~\ref{fig:phishssl} provides an overview of proposed method. The proposed PhishSSL comprises five sequential stages that collectively enable self-supervised phishing detection: (1) feature representation construction from multi-modal website data, (2) embedding encoding with adaptive feature attention, (3) contrastive view generation through sophisticated tabular augmentation, (4) triplet margin loss optimization for discriminative learning, and (5) inference-based classification through learned representations. Algorithm \ref{alg:phishssl} details the overall logic of each stage.

\begin{algorithm}[h]
\caption{Proposed PhishSSL Framework}
\label{alg:phishssl}
\footnotesize
\KwIn{Raw website data $(URL, HTML, Domain)$, margin $m$, augmentation config $\mathcal{A}$, training epochs $T$}
\KwOut{Trained embedding function $f$, phishing classifier $\mathcal{C}$}

\SetKwFunction{FRC}{FeatureConstruction}
\SetKwFunction{ENC}{EmbeddingEncoder}
\SetKwFunction{AUG}{AugmentationGenerator}
\SetKwFunction{CTL}{ContrastiveLearning}
\SetKwFunction{INF}{InferenceSetup}
\SetKwFunction{ATT}{FeatureAttention}

\SetKwProg{Fn}{Function}{:}{}
\Fn{\textsc{PhishSSL-Train}{}}{
    \tcp{Stage 1: Feature Representation Construction}
    \ForEach{website sample $w$}{
        $x_{url}, x_{html}, x_{ext} \leftarrow$ \FRC{$w.URL, w.HTML, w.Domain$}\;
        $x \leftarrow [x_{url}; x_{html}; x_{ext}] \in \mathbb{R}^{87}$\;
    }
    
    \tcp{Stage 2: Embedding Encoder with Feature Attention}
    $\theta_{att} \leftarrow$ Initialize attention parameters\;
    
    \For{epoch $t = 1$ to $T$}{
        \ForEach{batch $\mathcal{B}$}{
            \tcp{Stage 3: Contrastive View Generation via Tabular Augmentation}
            $\{x_{anchor}, x_{positive}\} \leftarrow$ \AUG{$\mathcal{B}, \mathcal{A}$}\;
            
            \tcp{Feature Attention and Embedding}
            $\alpha_{weights} \leftarrow$ \ATT{$x_{anchor}; \theta_{att}$}\;
            $z_{anchor} \leftarrow$ \ENC{$x_{anchor} \odot \alpha_{weights}$}\;
            $z_{positive} \leftarrow$ \ENC{$x_{positive} \odot \alpha_{weights}$}\;
            
            \tcp{Stage 4: Contrastive Learning with Triplet Margin Loss}
            $z_{negative} \leftarrow \text{shuffle}(z_{positive})$\;
            $\mathcal{L} \leftarrow$ \CTL{$z_{anchor}, z_{positive}, z_{negative}, m$}\;
            Update parameters via $\nabla \mathcal{L}$\;
        }
    }
    
    \tcp{Stage 5: Inference via Embedding Similarity}
        \tcp{Construct class prototypes used in the evaluation phase}
    $\boldsymbol{\mu}_{legitimate} \leftarrow \frac{1}{|S_{legitimate}|} \sum_{\boldsymbol{x}_i \in S_{legitimate}} f_\theta(\boldsymbol{x}_i)$\;
    $\boldsymbol{\mu}_{phishing} \leftarrow \frac{1}{|S_{phishing}|} \sum_{\boldsymbol{x}_i \in S_{phishing}} f_\theta(\boldsymbol{x}_i)$\;
    
    \tcp{Setup inference classifier using centroids}
    $\mathcal{C} \leftarrow$ \INF{$\boldsymbol{\mu}_{legitimate}, \boldsymbol{\mu}_{phishing}$}\;
}
\end{algorithm}

\subsection{Feature Representation Construction}\label{sec:feature}

Effective phishing detection requires comprehensive understanding of website characteristics across multiple dimensions. Therefore, our framework follows a systematic extraction of 87-dimensional feature vectors as in \cite{HANNOUSSE2021104347}, as shown in Table \ref{tab:features}, that capture complementary aspects of website behavior and structure.

The feature construction process transforms heterogeneous website data into standardized numerical representations suitable for machine learning. The standardization enables consistent comparison and learning across diverse website types and structures. The comprehensive feature vector is mathematically formulated as: $x = [x_{\text{url}}; x_{\text{html}}; x_{\text{ext}}] \in \mathbb{R}^{87}$, where the concatenation operator $[\cdot; \cdot; \cdot]$ combines three complementary feature domains: $x_{url} \in \mathbb{R}^{56}$ captures URL structural properties, $x_{html} \in \mathbb{R}^{24}$ represents HTML content characteristics, and $x_{ext} \in \mathbb{R}^{7}$ encodes external service indicators.

\begin{table}[h]
\centering
\caption{Comprehensive Feature Taxonomy for Phishing Detection}
\resizebox{\textwidth}{!}{
\begin{tabular}{|l|l|l|c|}
\hline
\textbf{Category} & \textbf{Feature Subcategory} & \textbf{Specific Features} & \textbf{Count} \\
\hline
\multirow{8}{*}{\begin{tabular}[c]{@{}l@{}}URL Structure\\ Features (56)\end{tabular}} 
& Length Metrics & url\_length, hostname\_length & 2 \\
& Character Frequency & nb\_dots, nb\_hyphens, nb\_at, nb\_qm, nb\_and, nb\_or, & 20 \\
& & nb\_eq, nb\_underscore, nb\_tilde, nb\_percent, nb\_slash, & \\
& & nb\_star, nb\_colon, nb\_comma, nb\_semicolon, nb\_dollar, & \\
& & nb\_space, nb\_www, nb\_com, nb\_dslash & \\
& Security Indicators & ip, https\_token, punycode, port & 4 \\
& Structural Properties & tld\_in\_path, tld\_in\_subdomain, abnormal\_subdomain, & 5 \\
& & nb\_subdomains, prefix\_suffix & \\
& Content Analysis & http\_in\_path, ratio\_digits\_url, ratio\_digits\_host & 3 \\
& Redirection Patterns & nb\_redirection, nb\_external\_redirection & 2 \\
& Lexical Properties & length\_words\_raw, char\_repeat, shortest\_words\_raw, & 12 \\
& & shortest\_word\_host, shortest\_word\_path, longest\_words\_raw, & \\
& & longest\_word\_host, longest\_word\_path, avg\_words\_raw, & \\
& & avg\_word\_host, avg\_word\_path, random\_domain & \\
& Deception Indicators & phish\_hints, domain\_in\_brand, brand\_in\_subdomain, & 5 \\
& & brand\_in\_path, shortening\_service & \\
& External Validation & path\_extension, suspicious\_tld, statistical\_report & 3 \\
\hline
\multirow{4}{*}{\begin{tabular}[c]{@{}l@{}}HTML Content\\ Features (24)\end{tabular}}
& Hyperlink Analysis & nb\_hyperlinks, ratio\_intHyperlinks, ratio\_extHyperlinks, & 4 \\
& & ratio\_nullHyperlinks & \\
& Resource Inclusion & nb\_extCSS, external\_favicon, links\_in\_tags, & 6 \\
& & ratio\_intMedia, ratio\_extMedia, ratio\_intRedirection & \\
& Form Characteristics & login\_form, submit\_email, sfh & 3 \\
& Behavioral Indicators & iframe, popup\_window, safe\_anchor, onmouseover, & 11 \\
& & right\_click, empty\_title, domain\_in\_title, & \\
& & domain\_with\_copyright, ratio\_extRedirection, & \\
& & ratio\_intErrors, ratio\_extErrors & \\
\hline
\multirow{3}{*}{\begin{tabular}[c]{@{}l@{}}External Service\\ Features (7)\end{tabular}}
& Domain Registration & whois\_registered\_domain, domain\_registration\_length, & 3 \\
& & domain\_age & \\
& Web Presence & web\_traffic, dns\_record, google\_index & 3 \\
& Authority Metrics & page\_rank & 1 \\
\hline
\end{tabular}\label{tab:features}
}
\end{table}

\textbf{URL Structural Features} capture lexical and syntactic properties that often reveal deceptive intent. These include length distributions (unusually long URLs often indicate obfuscation), character frequency patterns (excessive special characters suggest manipulation), subdomain configurations (suspicious subdomain structures), and redirection analysis (multiple redirections often hide true destinations). The ratio of digits in URLs and hostnames provides insights into automated generation patterns common in phishing attacks.

\textbf{HTML Content Features} quantify webpage composition and behavioral characteristics. Hyperlink density ratios reveal external dependency patterns, while resource inclusion analysis identifies suspicious external content loading. Form characteristics are particularly important as phishing sites often contain deceptive login forms. Behavioral indicators such as iframe usage, popup windows, and JavaScript-based redirection provide additional discrimination signals.

\textbf{External Service Features} leverage third-party data sources to assess domain legitimacy and reputation. WHOIS registration data reveals domain age and registration patterns, while DNS record analysis confirms proper domain configuration. Web traffic statistics and search engine indexing status provide insights into domain popularity and legitimacy that are difficult to fabricate quickly.

\subsection{Embedding Encoder with Feature Attention}

Traditional neural networks treat all input features equally, potentially diluting the impact of highly discriminative characteristics. In phishing detection, certain features (such as URL length, suspicious TLD usage, or external redirections) may be more indicative of malicious intent than others. Furthermore, the relative importance of features may vary across different types of websites and attack strategies.

To address these challenges, our framework incorporates a learnable feature attention mechanism that dynamically weights input features based on their discriminative power for the specific task. This adaptive approach allows the model to focus on the most relevant characteristics while maintaining the ability to leverage the full feature set when beneficial.

The feature attention mechanism employs a two-layer neural network to compute element-wise importance scores: $\alpha = \sigma(W_2 \cdot \text{ReLU}(W_1 \cdot x + b_1) + b_2)$, where $W_1 \in \mathbb{R}^{d_{att} \times 87}$ and $W_2 \in \mathbb{R}^{87 \times d_{att}}$ represent learnable transformation matrices, $b_1 \in \mathbb{R}^{d_{att}}$ and $b_2 \in \mathbb{R}^{87}$ are bias vectors, $d_{att}$ denotes the attention hidden dimension, and $\sigma(\cdot)$ represents the sigmoid activation function. The sigmoid ensures attention weights $\alpha \in [0,1]^{87}$, providing interpretable importance scores for each feature.

The attended feature representation combines original features with learned importance weights: $x_{\text{att}} = x \odot \alpha + \beta \cdot x$, where $\odot$ denotes element-wise multiplication, and $\beta$ is a residual connection parameter that prevents complete feature suppression. This formulation ensures that even features with low attention weights retain some influence, maintaining model robustness.

The embedding encoder transforms attended features through a multi-layer architecture:
\begin{equation}
\begin{aligned}
h_1 &= \text{ReLU}(W_{enc1} \cdot x_{att} + b_{enc1}) \\
h_2 &= \text{Dropout}(\text{ReLU}(W_{enc2} \cdot h_1 + b_{enc2})) \\
z_{raw} &= W_{enc3} \cdot h_2 + b_{enc3}, \qquad
z = \frac{z_{raw}}{\|z_{raw}\|_2}
\end{aligned}
\end{equation}

where $W_{enc1} \in \mathbb{R}^{d_1 \times 87}$, $W_{enc2} \in \mathbb{R}^{d_2 \times d_1}$, and $W_{enc3} \in \mathbb{R}^{d \times d_2}$ are encoding layer weights, and $z \in \mathbb{R}^d$ represents the final L2-normalized embedding.

Layer normalization is applied after each transformation to ensure training stability: $\text{LayerNorm}(h) = \gamma \cdot \frac{h - \mu}{\sqrt{\sigma^2 + \epsilon}} + \beta$, where $\mu$ and $\sigma^2$ are the mean and variance of layer activations, $\gamma$ and $\beta$ are learnable scale and shift parameters, and $\epsilon$ is a small constant for numerical stability.

The L2 normalization of final embeddings is crucial for contrastive learning as it ensures that distance computations focus on angular relationships rather than magnitude differences: \( d(z_i, z_j) = \|z_i - z_j\|_2 = \sqrt{2 - 2 \langle z_i, z_j \rangle} \).

This normalization enables the model to learn discriminative representations where legitimate and phishing websites cluster in distinct regions of the unit hypersphere.

\subsection{Contrastive View Generation via Tabular Augmentation}\label{sec:data_aug}

Contrastive learning requires the generation of semantically consistent positive pairs while maintaining sufficient diversity to drive meaningful representation learning. Our framework adopts two complementary augmentation strategies designed specifically for tabular phishing data: dropout-based augmentation and traditional tabular augmentation techniques. These approaches generate positive pairs while introducing controlled perturbations essential for robust representation learning.

\textbf{Dropout-based Augmentation} This approach leverages the stochastic nature of dropout to generate different views of the same input during training. By applying distinct dropout masks to identical inputs, the model observes varied but semantically equivalent representations, encouraging learning of robust features that remain consistent across different network states. The dropout augmentation process generates positive pairs through: \( z_{\text{anchor}} = f(x; \text{mask}_1), \quad z_{\text{positive}} = f(x; \text{mask}_2) \), where $\text{mask}_1$ and $\text{mask}_2$ represent independent Bernoulli random variables: \( \text{mask}_i[j] \sim \text{Bernoulli}(1 - p_{\text{dropout}}) \), with $p_{dropout}$ controlling augmentation intensity. Higher dropout rates increase view diversity but may compromise semantic preservation.

\textbf{Traditional Tabular Augmentation} This approach combines multiple transformation techniques specifically adapted for tabular data:

1. CutMix Augmentation: Randomly mixes features between different samples to create synthetic training examples: \( x_{\text{cutmix}} = \lambda \cdot x + (1 - \lambda) \cdot x_j \), where $\lambda \sim \text{Beta}(\alpha, \alpha)$ controls mixing intensity, and $x_j$ is randomly selected from the batch. The beta distribution ensures most mixing occurs near the extremes (0 or 1), preserving semantic content.

2. Gaussian Noise Injection: Adds controlled noise to numerical features: \( x_{\text{noisy}} = x + \epsilon, \quad \epsilon \sim \mathcal{N}(0, \sigma^2 I) \), where $\sigma^2$ controls noise magnitude. The noise level is carefully calibrated to avoid overwhelming feature signals while providing sufficient perturbation for contrastive learning.

3. Feature Corruption: To simulate measurement noise and enhance robustness, we randomly corrupt a subset of input features. Specifically, each feature \( x[i] \) is retained with probability \( 1 - p_{\text{corrupt}} \), and replaced with a random value with probability \( p_{\text{corrupt}} \), i.e., \( x_{\text{corrupt}}[i] = x[i] \) with probability \( 1 - p_{\text{corrupt}} \), otherwise \( x_{\text{corrupt}}[i] = \text{random}() \).

\subsection{Contrastive Learning with Triplet Margin Loss}

The core learning mechanism of our framework operates through triplet margin loss optimization, which shapes the embedding space geometry to achieve optimal separation between legitimate and phishing websites. Unlike traditional classification approaches that learn decision boundaries, contrastive learning focuses on learning meaningful distance relationships in the representation space.

The triplet formulation addresses a fundamental challenge in unsupervised learning: how to learn discriminative representations without explicit class labels. By leveraging the natural assumption that augmented views of the same website should be more similar than different websites, the framework learns to encode semantic relationships into the embedding space.

\textbf{Triplet Construction and Semantics:} Each training triplet consists of three components:
- \textbf{Anchor} ($z_a$): Original sample embedding
- \textbf{Positive} ($z_p$): Augmented view of the same sample  
- \textbf{Negative} ($z_n$): Embedding from a different sample.

The triplet margin loss enforces the constraint that positive pairs should be closer than negative pairs by at least a margin $m$:

\begin{equation}
\mathcal{L}_{triplet} = \max(0, d(z_a, z_p) - d(z_a, z_n) + m)
\end{equation}

where $d(\cdot, \cdot)$ represents the L2 distance between normalized embeddings:

\begin{equation}
d(z_i, z_j) = \|z_i - z_j\|_2 = \sqrt{\sum_{k=1}^{d}(z_i^{(k)} - z_j^{(k)})^2}
\end{equation}

\textbf{Positive Sample Generation:} Positive samples $(z_a, z_p)$ are generated with the augmentation strategies described in Section \ref{sec:data_aug}. The key insight is that different augmented views of the same website should produce similar embeddings, regardless of the specific augmentation applied. This consistency requirement forces the model to learn invariant representations that capture essential website characteristics rather than superficial variations. Mathematically, positive pairs satisfy the semantic equivalence constraint: $\text{sem}(x) = \text{sem}(\mathcal{T}(x, \theta))$, where $\text{sem}(\cdot)$ represents semantic content and $\mathcal{T}(\cdot, \theta)$ denotes augmentation transformation.

\textbf{Negative Sample Generation} Negative samples are generated through in-batch random shuffling, ensuring computational efficiency while maintaining learning effectiveness: $z_n^{(i)} = z_p^{(\pi(i))}$, where $\pi: \{1, 2, \ldots, N\} \rightarrow \{1, 2, \ldots, N\}$ represents a random permutation satisfying $\pi(i) \neq i$ for all $i$. This constraint prevents trivial negative pairs where a sample might be paired with itself.

\textbf{Margin Parameter:} The margin parameter $m$ plays a crucial role in determining embedding space geometry. A larger margin enforces stronger separation between positive and negative pairs, potentially leading to more discriminative representations but risking overfitting. The optimal margin balances separation strength with generalization capability: $m^* = \arg\min_m \mathbb{E}[\mathcal{L}_{\text{validation}}(m)]$.

\textbf{Batch-wise Loss Computation:} The complete training objective aggregates triplet losses across all samples in a batch:

\begin{equation}
\mathcal{L}_{batch} = \frac{1}{N} \sum_{i=1}^{N} \max(0, \|z_a^{(i)} - z_p^{(i)}\|_2 - \|z_a^{(i)} - z_n^{(i)}\|_2 + m)
\end{equation}

where $N$ represents batch size. This formulation enables efficient parallel computation while maintaining the semantic relationships essential for effective contrastive learning.

\textbf{Embedding Space Properties:} The triplet loss optimization results in an embedding space with desirable geometric properties: 1. Cluster Formation: Legitimate and phishing websites naturally cluster in distinct regions; 2. Intra-class Cohesion: Similar websites (same class) maintain small pairwise distances; 3. Inter-class Separation: Different website types exhibit large pairwise distances; 4. Metric Properties: The learned distance function satisfies triangle inequality and symmetry. These properties enable effective downstream classification through simple distance-based methods, eliminating the need for complex classifiers and improving interpretability of the learned representations.

\textbf{Training Dynamics and Convergence:} The contrastive learning process exhibits characteristic training dynamics where the model initially learns coarse-grained distinctions before refining fine-grained patterns. The triplet loss typically decreases rapidly in early epochs as the model learns basic clustering, then gradually refines boundaries through continued optimization.

Convergence is achieved when the embedding space reaches a stable configuration where:

\begin{equation}
\lim_{t \rightarrow \infty} \mathbb{E}[\mathcal{L}_{triplet}(t)] = \mathbb{E}[\max(0, d_{positive} - d_{negative} + m)]
\end{equation}

approaches a minimum value determined by the inherent separability of the dataset and the chosen margin parameter.

\subsection{Inference via Embedding Similarity}

During inference, we adopt a prototype-based evaluation strategy. Each test embedding is compared against class-specific prototypes, and classification is determined by nearest-prototype similarity. Training remains fully label-free, and labels are only involved in the evaluation.

\textbf{Centroid Construction and Similarity Computation:} Given a class $c \in {\text{legitimate}, \text{phishing}}$, its corresponding centroid $\boldsymbol{\mu}_c$ is defined as the mean of reference embeddings used in the evaluation protocol: $\boldsymbol{\mu}_c = \frac{1}{|S_c|} \sum_{\boldsymbol{x}_i \in S_c} f_\theta(\boldsymbol{x}_i)$. Here, $S_c$ denotes the set of reference samples for class $c$ used solely in evaluation, $f_\theta(\cdot)$ represents the trained encoder, and $d_c$ is the Euclidean distance between the test embedding $\boldsymbol{z}_{\text{test}}$ and the prototype. Alternatively, cosine similarity captures the angular relationship between the test embedding and the centroid: \( d_c = \|\boldsymbol{z}_{\text{test}} - \boldsymbol{\mu}_c\|_2 \) and \( s_c = \frac{\boldsymbol{z}_{\text{test}} \cdot \boldsymbol{\mu}_c}{\|\boldsymbol{z}_{\text{test}}\|_2 \|\boldsymbol{\mu}_c\|_2} \).

\textbf{Probabilistic Classification and Decision Making:} For Euclidean-based inference, the distance is transformed into class probabilities using a softmax-like inverse exponential function. For cosine similarity-based inference, a sigmoid activation function is employed to convert similarities into probability estimates:

\begin{equation}
P(y = c | \boldsymbol{z}_{\text{test}}) = \frac{\exp(-d_c)}{\sum_{c'} \exp(-d_{c'})}, \quad P(y = c | \boldsymbol{z}_{\text{test}}) = \sigma(s_c)
\end{equation}

In both cases, the final classification decision is determined by comparing predicted class probabilities against a predefined threshold.

\section{Evaluation}\label{eval}
\subsection{Dataset \& Testbed}
We conduct our experiments on three phishing detection datasets. The primary Benchmark dataset contains 11,430 samples with 87 extracted features, evenly split between phishing and legitimate instances \cite{HANNOUSSE2021104347}. To evaluate the robustness of PhishSSL, we further test it on two additional datasets: the Tan dataset \cite{tan2018phishing}, with 10,000 samples (5,000 phishing and 5,000 legitimate) and 48 features, and the Grega dataset \cite{vrbancic2020phishing}, with 20,000 samples (10,000 phishing and 10,000 legitimate) and 111 features. The Benchmark dataset \cite{HANNOUSSE2021104347} provides a balanced mix of URL structural, HTML content-based, and external service features. The Tan dataset \cite{tan2018phishing} emphasizes compact structural and behavioral cues while omitting most external signals. The Grega dataset \cite{vrbancic2020phishing} concentrates on fine-grained URL and protocol-level properties but includes relatively few content features. Using these datasets with diverse feature compositions allows us to examine the effectiveness of the proposed method under varying feature settings. All datasets adopt a consistent 60\%/20\%/20\% Train/Validation/Test split. Training remains fully unsupervised, with labels used solely in the evaluation protocol to derive class prototypes and for validation and testing of model performance.

All models are trained with a batch size of 128 and a learning rate of 1e-3. PhishSSL uses a 256-dimensional encoder, a 128-dimensional projection head, and a 64-dimensional attention module. Baselines share the same settings for fair comparison. Each model is run for 20 epochs, and the best ROC AUC is reported. Experiments are conducted on an H20 NVLink GPU server with 96 GB of memory.
\subsection{Baselines}
To evaluate the effectiveness of the proposed PhishSSL framework, four representative unsupervised and self-supervised baseline models are employed for comparative analysis. \textbf{K-Means Clustering} partitions the feature space into distinct clusters based on Euclidean distance minimization, assuming that legitimate and phishing websites form natural clusters. Classification is performed by assigning test samples to the nearest cluster centroid. \textbf{Isolation Forest} employs an anomaly detection approach based on the isolation principle, where anomalous samples are easier to isolate than normal samples. The algorithm constructs an ensemble of random trees with anomaly scores determined by the average path length required to isolate each sample. \textbf{Autoencoder} implements a neural network-based reconstruction approach consisting of an encoder-decoder architecture that learns to compress input features into a latent representation and reconstruct the original input. Reconstruction error serves as the anomaly score. \textbf{Variational Autoencoder (VAE)} extends the standard autoencoder by introducing probabilistic encoding through variational inference. The model learns probability distributions over the latent space, incorporating Kullback-Leibler divergence regularization alongside reconstruction loss for more robust anomaly detection.
\subsection{Evaluation Metrics}
The performance of all models is assessed using a comprehensive set of classification metrics that provide insights into different aspects of phishing detection effectiveness. \textbf{Confusion Matrix} provides a detailed breakdown of classification results by tabulating true positives (TP), true negatives (TN), false positives (FP), and false negatives (FN), enabling a thorough analysis across both legitimate and phishing categories.

\textbf{Accuracy} measures the overall proportion of correctly classified samples and is defined as $\text{Accuracy} = \frac{TP + TN}{TP + TN + FP + FN}$. \textbf{Precision}, defined as $\text{Precision} = \frac{TP}{TP + FP}$, quantifies the proportion of predicted phishing websites that are actually malicious, indicating the model's ability to avoid false alarms. \textbf{Recall} measures the proportion of actual phishing websites correctly identified and is given by $\text{Recall} = \frac{TP}{TP + FN}$, reflecting the model's capability to detect malicious samples.

\textbf{F1-Score} provides a harmonic mean of precision and recall, defined as $\text{F1} = 2 \cdot \frac{\text{Precision} \times \text{Recall}}{\text{Precision} + \text{Recall}}$. Finally, \textbf{ROC AUC} evaluates the model’s discriminative ability across all classification thresholds by plotting the true positive rate against the false positive rate, with the area under the curve ranging from 0 to 1—higher values indicate better threshold-independent performance.

\section{Results}\label{results}
\subsection{Comparative Analysis}
\begin{figure*}[!ht]
    \centering
    \begin{subfigure}[b]{0.30\linewidth}
        \includegraphics[width=\linewidth, height=3.5cm]{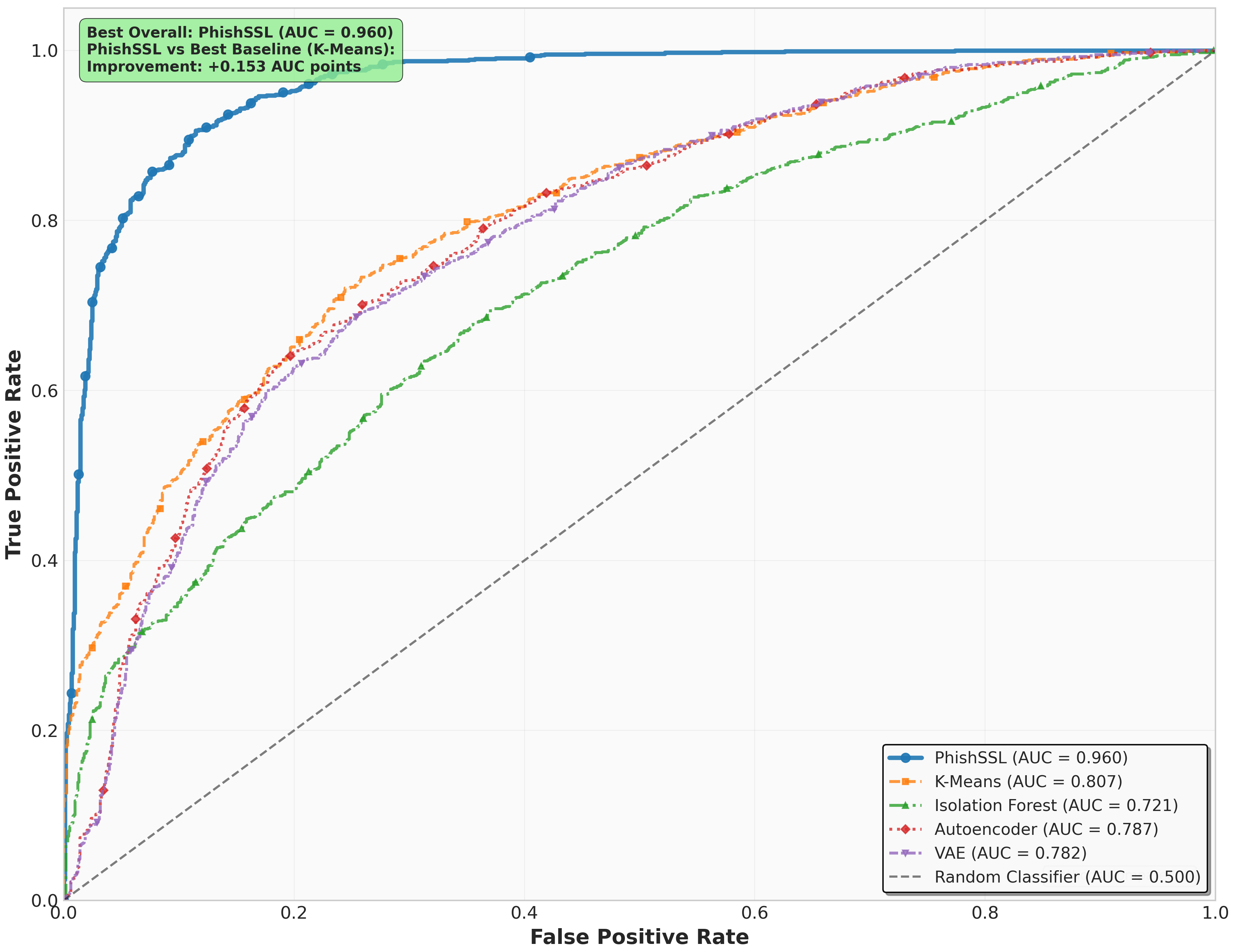}
        \caption{Benchmark dataset}
        \label{fig:roc_comparison_m}
    \end{subfigure}
    \hfill
    \begin{subfigure}[b]{0.30\linewidth}
        \includegraphics[width=\linewidth, height=3.5cm]{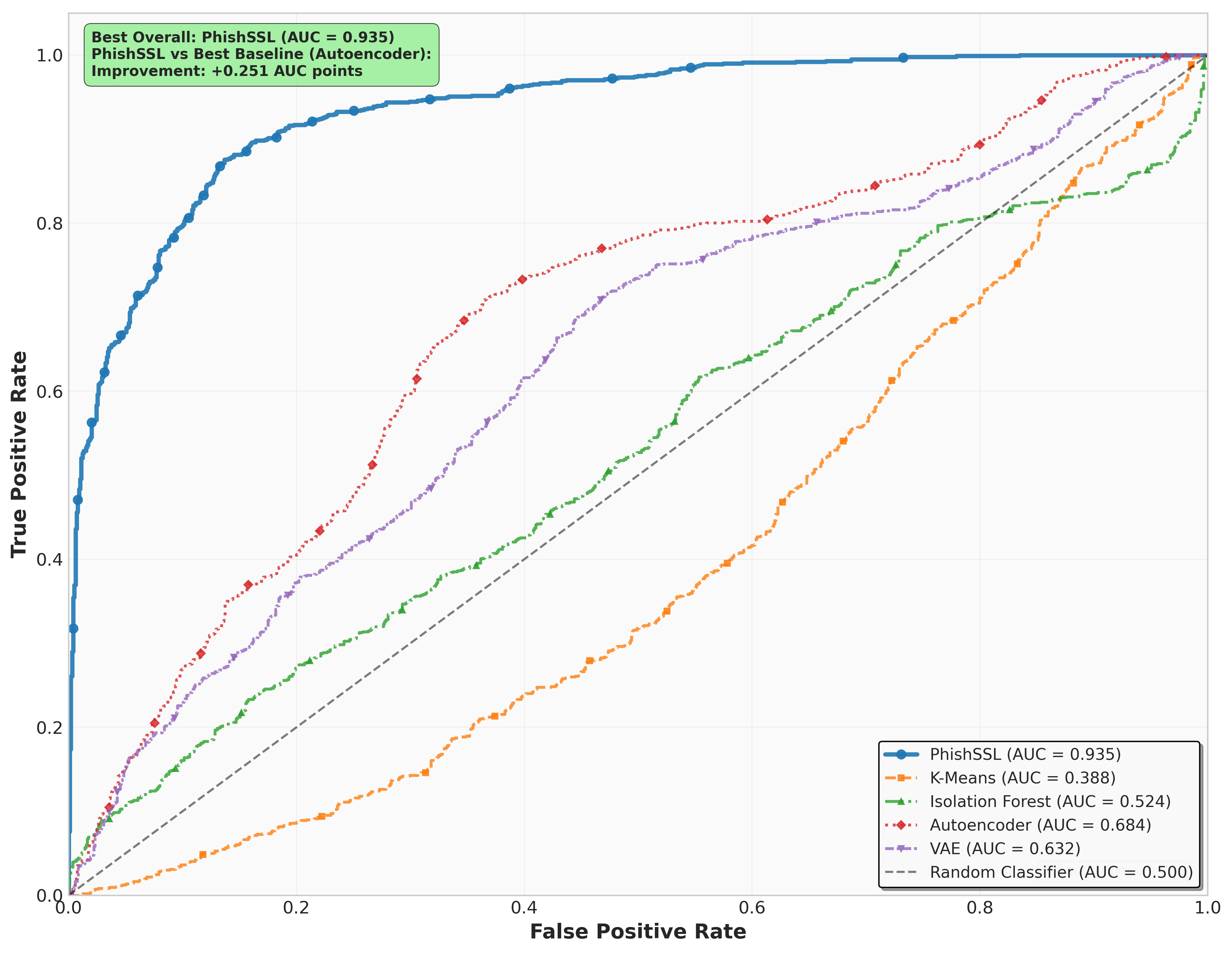}
        \caption{Tan dataset}
        \label{fig:roc_comparison_t}
    \end{subfigure}
    \hfill
    \begin{subfigure}[b]{0.30\linewidth}
        \includegraphics[width=\linewidth, height=3.5cm]{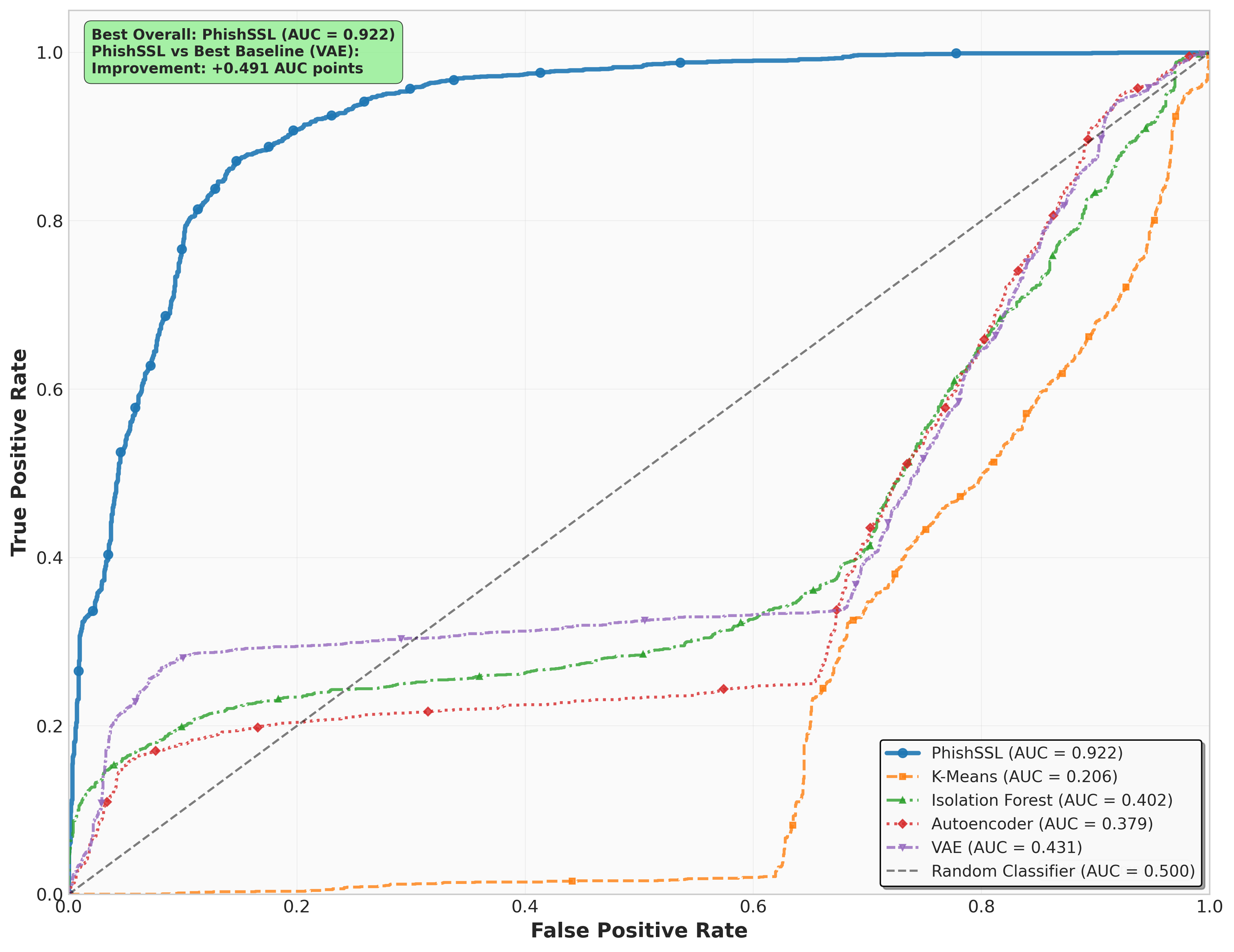}
        \caption{Grega dataset}
        \label{fig:roc_comparison_g}
    \end{subfigure}

    \caption{Comparison of ROC curves across different detection models on three phishing datasets.}
    \label{fig:model_analysis}
\end{figure*}

\begin{figure*}[!ht]
    \centering

    \begin{subfigure}[b]{0.30\linewidth}
        \includegraphics[width=\linewidth, height=3.5cm]{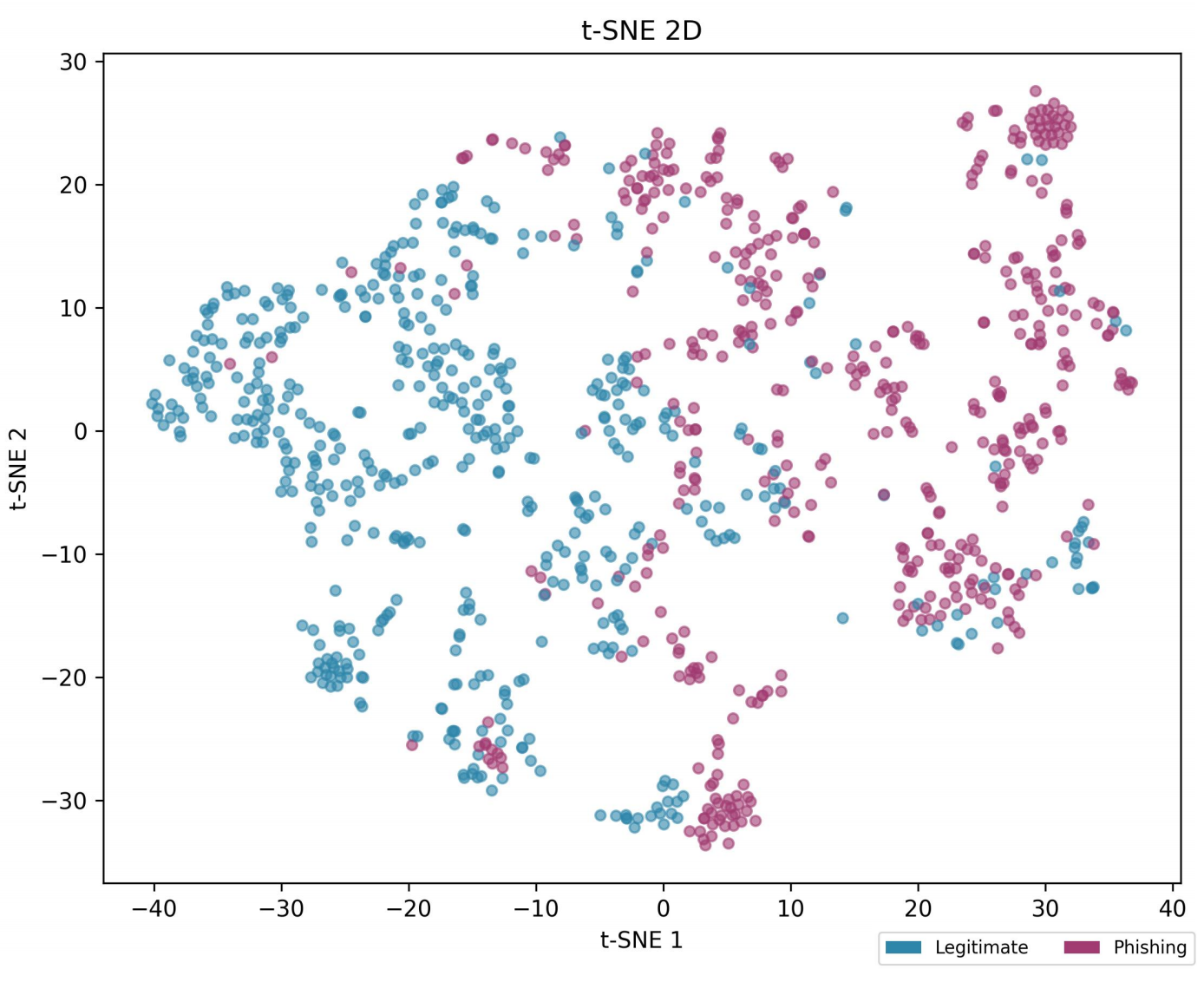}
        \caption{Benchmark dataset}
        \label{fig:t_sne_m}
    \end{subfigure}
    \hfill
    \begin{subfigure}[b]{0.30\linewidth}
        \includegraphics[width=\linewidth, height=3.5cm]{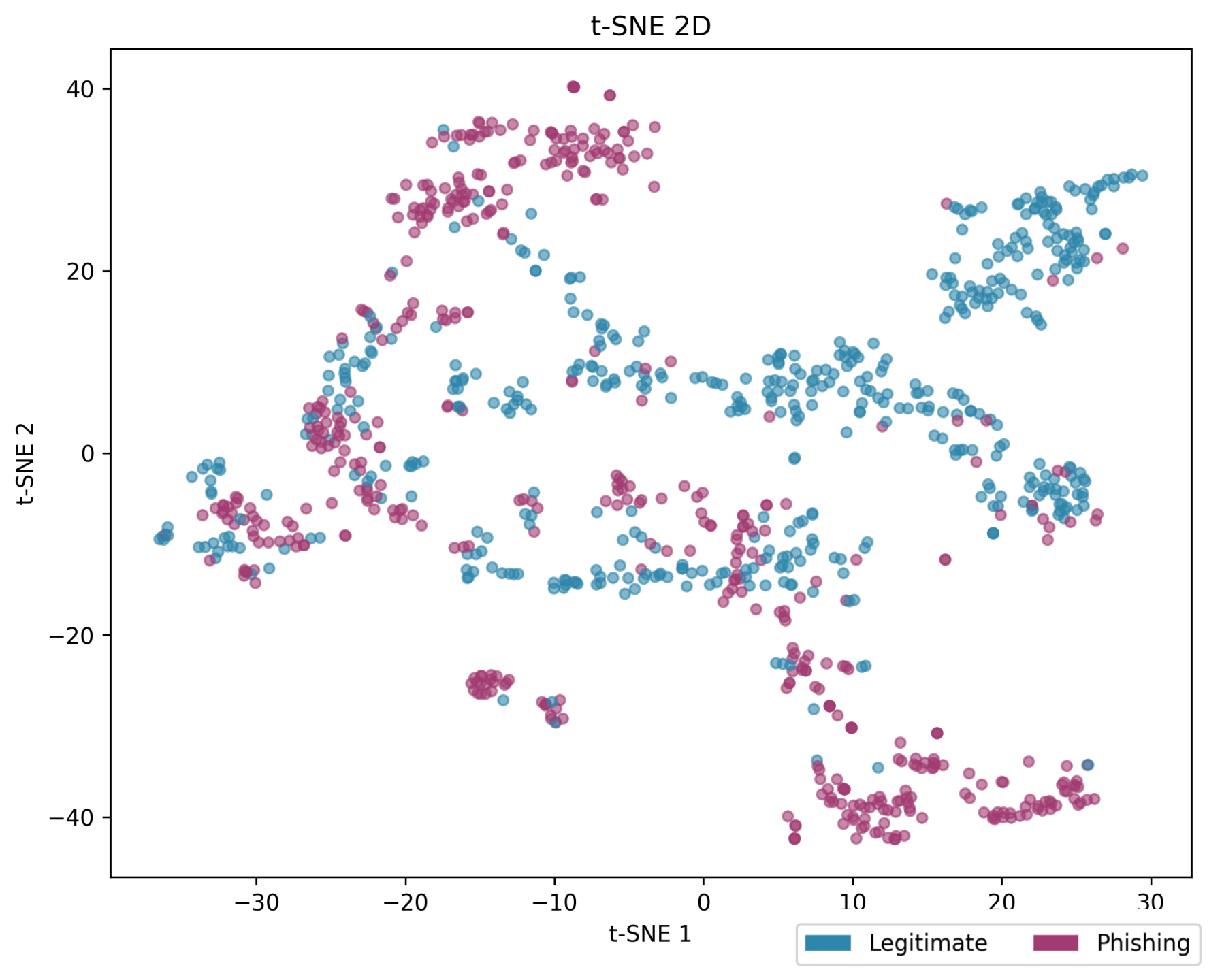}
        \caption{Tan dataset}
        \label{fig:t_sne_t}
    \end{subfigure}
    \hfill
    \begin{subfigure}[b]{0.30\linewidth}
        \includegraphics[width=\linewidth, height=3.5cm]{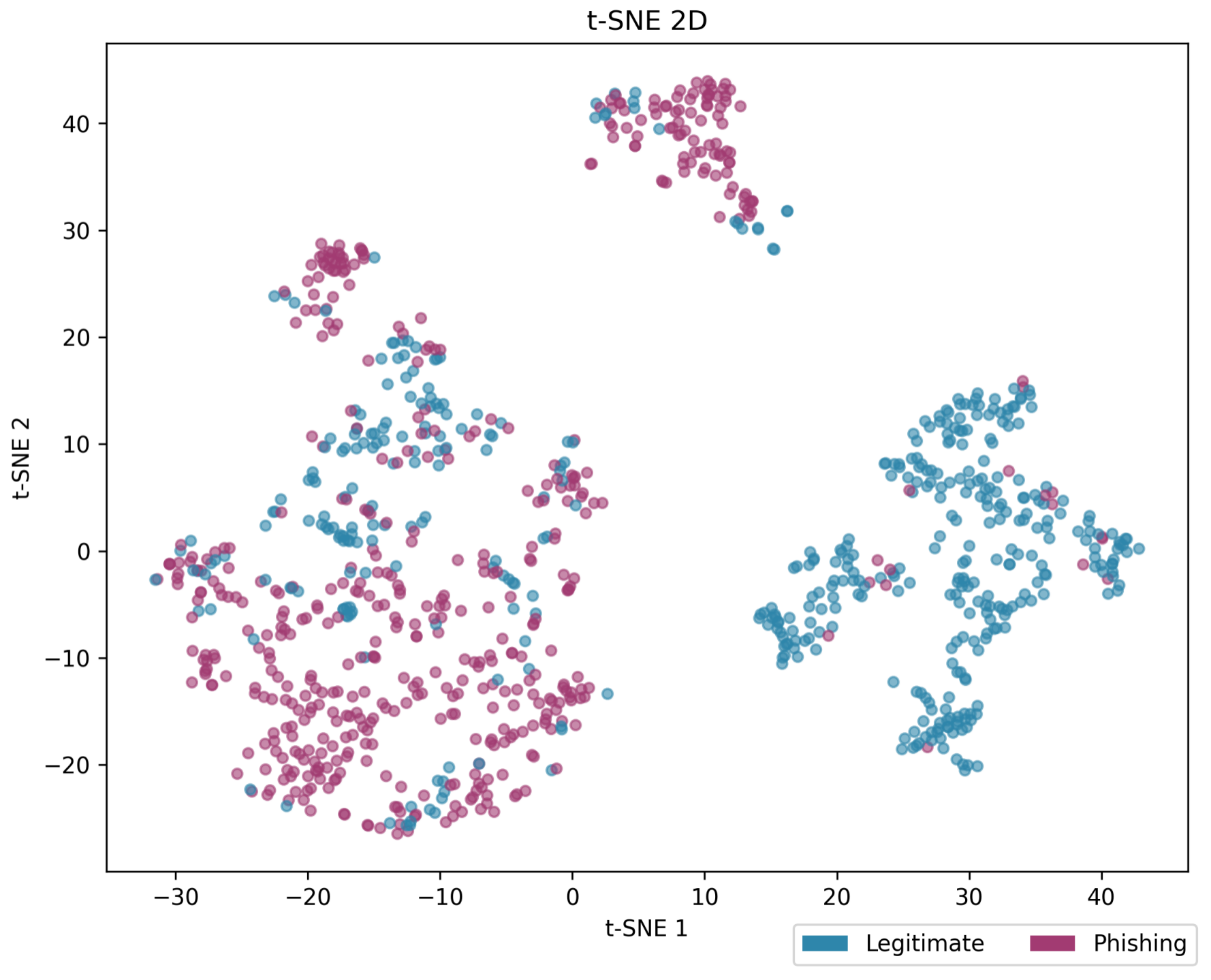}
        \caption{Grega dataset}
        \label{fig:t_sne_g}
    \end{subfigure}

    \caption{t-SNE visualization of the learned embeddings on three phishing datasets.}
    \label{fig:t_sne}
\end{figure*}

To demonstrate the performance and robustness of the proposed PhishSSL framework, we report results across three phishing detection datasets with distinct feature sets. Table~\ref{tab:anomaly_models_metrics_split_cm_front}, Fig.~\ref{fig:model_analysis}, and Fig.~\ref{fig:t_sne} present a comprehensive comparison against four baseline models using standard evaluation metrics and visual diagnostics.

\begin{table}[htbp]
\centering
\caption{Performance and Confusion Matrix of Models on Phishing Datasets.}
\label{tab:anomaly_models_metrics_split_cm_front}
\setlength{\tabcolsep}{3pt}
\begin{tabular}{@{}lcccccccc@{}}
\toprule
\textbf{Dataset / Model} & \textbf{TN} & \textbf{FP} & \textbf{FN} & \textbf{TP} & \textbf{Acc.} & \textbf{Prec.} & \textbf{Rec.} & \textbf{F1} \\
\midrule
\multicolumn{9}{l}{\textbf{Benchmark Dataset \cite{HANNOUSSE2021104347}}} \\
K-Means            & 1127 & 7   & 906 & 246 & 0.6006 & \textbf{0.9723} & 0.2135 & 0.3502 \\
Isolation Forest   & 1114 & 20  & 959 & 193 & 0.5717 & 0.9061 & 0.1675 & 0.2828 \\
Autoencoder        & 998  & 136 & 579 & 573 & 0.6872 & 0.8082 & 0.4974 & 0.6158 \\
VAE                & 1000 & 134 & 607 & 545 & 0.6759 & 0.8027 & 0.4731 & 0.5953 \\
\textbf{PhishSSL}  & 1037 & 134 & 105 & 1010 & \textbf{0.8955} & 0.8829 & \textbf{0.9058} & \textbf{0.8942} \\
\midrule
\multicolumn{9}{l}{\textbf{Tan Dataset \cite{tan2018phishing}}} \\
K-Means            & 591  & 443 & 725 & 241 & 0.4160 & 0.3523 & 0.2495 & 0.2921 \\
Isolation Forest   & 968  & 66  & 854 & 112 & 0.5400 & 0.6292 & 0.1159 & 0.1958 \\
Autoencoder        & 821  & 213 & 562 & 404 & 0.6125 & 0.6548 & 0.4182 & 0.5104 \\
VAE                & 790  & 244 & 581 & 385 & 0.5875 & 0.6121 & 0.3986 & 0.4828 \\
\textbf{PhishSSL}  & 852 & 135 & 126 & 887 & \textbf{0.8695} & \textbf{0.8679} & \textbf{0.8756} & \textbf{0.8717} \\
\midrule
\multicolumn{9}{l}{\textbf{Grega Dataset \cite{vrbancic2020phishing}}} \\
K-Means            & 760  & 1237 & 1961 & 42  & 0.2005 & 0.0328 & 0.0210 & 0.0256 \\
Isolation Forest   & 1921 & 76   & 1697 & 306 & 0.5567 & 0.8010 & 0.1528 & 0.2566 \\
Autoencoder        & 1677 & 320  & 1609 & 394 & 0.5178 & 0.5518 & 0.1967 & 0.2900 \\
VAE                & 1775 & 222  & 1430 & 573 & 0.5870 & 0.7208 & 0.2861 & 0.4096 \\
\textbf{PhishSSL}  & 1704 & 293 & 258 & 1745 & \textbf{0.8622} & \textbf{0.8562} & \textbf{0.8712} & \textbf{0.8636} \\
\bottomrule
\end{tabular}
\end{table}

On the Benchmark dataset, PhishSSL achieves an ROC AUC of 0.960 and F1-score of 0.8942, outperforming the best baseline (K-Means) by a margin of +0.153 AUC points (Fig.~\ref{fig:roc_comparison_m}). Notably, while K-Means achieves high precision (0.9723), it suffers from extremely low recall (0.2135), suggesting it misses a large number of phishing websites. In contrast, PhishSSL maintains a balanced detection ability with a recall of 0.9058 and low false negative rate (FN=105), demonstrating its ability to identify most phishing attempts.

On the Tan dataset, PhishSSL remains dominant with an ROC AUC of 0.935, significantly exceeding the best-performing baseline (Autoencoder, AUC = 0.684) by +0.251 AUC points (Fig.~\ref{fig:roc_comparison_t}). It achieves an F1-score of 0.8717, maintaining both high precision (0.8679) and recall (0.8756), while other baselines exhibit severe drops in recall, with Isolation Forest falling to 0.1159 and K-Means to 0.2495, highlighting their poor generalization on cross-dataset phishing features.

The robustness of PhishSSL is further confirmed on the Grega dataset, where it achieves an ROC AUC of 0.922 and F1-score of 0.8636. Compared to the best baseline (VAE, AUC = 0.431), this represents an improvement of +0.491 AUC points (Fig.~\ref{fig:roc_comparison_g}). Other models suffer from high false positives and extremely poor recall, especially K-Means (recall = 0.0210) and Autoencoder (recall = 0.1967).

The t-SNE visualizations in Fig.~\ref{fig:t_sne} highlight the quality of the embedding space learned by PhishSSL. On all datasets, phishing and legitimate samples form well-separated clusters, especially on the Grega and Benchmark datasets, where PhishSSL distinctly isolates phishing clusters. This confirms the effectiveness of our contrastive learning mechanism combined with adaptive attention and tabular augmentation in generating discriminative and transferable representations.

Although the three datasets differ in their feature compositions, PhishSSL consistently achieves high and stable performance across all evaluation metrics, demonstrating strong robustness and adaptability for real-world phishing detection. We also observe that both the baselines and PhishSSL obtain higher performance on the Benchmark dataset than on Tan and Grega, likely because the Benchmark dataset provides a more comprehensive combination of features, offering richer and more complementary signals.

\subsection{Ablation Analysis}
To assess the contribution of individual components within the PhishSSL framework, we conduct an ablation study by selectively removing core modules, including feature attention, traditional tabular augmentation, and dropout-based augmentation. The results are summarized in Table~\ref{tab:phishssl_ablation}.

\begin{table}[htbp]
\centering
\setlength{\belowcaptionskip}{-0.2cm}
\caption{Ablation Analysis of PhishSSL on Benchmark dataset.}
\label{tab:phishssl_ablation}
\setlength{\tabcolsep}{1pt} % Adjust column spacing
\begin{tabular}{@{}lccccc@{}}
\toprule
\textbf{Configuration} & 
\textbf{ROC} & 
\textbf{Accuracy} & 
\textbf{Precision} & 
\textbf{Recall} & 
\textbf{F1-Score} \\
\midrule
\textbf{Baseline (All Components)}      & \textbf{0.9601} & \textbf{0.8955} & 0.8829 & \textbf{0.9058} & \textbf{0.8942} \\
\midrule
w/o Feature Attention           & 0.9491 & 0.8920 & \textbf{0.9049} & 0.8700 & 0.8871 \\
w/o Traditional Augmentation    & 0.9440 & 0.8780 & 0.8842 & 0.8628 & 0.8734 \\
w/o Dropout Augmentation        & 0.9169 & 0.8530 & 0.8338 & 0.8726 & 0.8528 \\
\bottomrule
\end{tabular}
\end{table}

The full PhishSSL model achieves the best overall performance across all metrics, indicating the synergistic effect of its design components. Removing the feature attention module results in a slight drop in ROC AUC (from 0.9601 to 0.9491), accompanied by a reduction in F1-score and recall, highlighting its role in emphasizing discriminative features and improving generalization. Eliminating traditional augmentation techniques (CutMix, Gaussian noise, and feature corruption) leads to a further decrease in performance, with ROC AUC dropping to 0.9440 and F1-score to 0.8734. This underscores the importance of diverse perturbations for encouraging invariant feature learning.

The most significant degradation occurs when dropout-based augmentation is disabled. Without stochastic dropout views, ROC AUC falls to 0.9169, and F1-score declines to 0.8528. These results confirm that dropout augmentation plays a central role in generating semantically consistent views for contrastive learning and in enhancing the robustness of the learned embeddings.

Overall, the ablation analysis demonstrates that each component in PhishSSL contributes meaningfully to the final performance, and their integration yields a robust and highly discriminative phishing detection system.

\section{Conclusion}\label{conclusion}
This study presented PhishSSL, a self-supervised contrastive learning framework for phishing website detection that avoids reliance on labeled training data. Extensive experiments across diverse datasets confirmed its robustness, generalization, and superiority over baseline methods. Looking forward, future directions include examining the impact of different supervised classification heads, incorporating multi-modal fusion by integrating heterogeneous features through cross-modal contrastive learning, and enhancing adversarial robustness to resist evolving evasion tactics. These avenues will further strengthen the effectiveness and reliability of phishing defenses in increasingly dynamic and adversarial Web environments.

%
% ---- Bibliography ----
%
% BibTeX users should specify bibliography style 'splncs04'.
% References will then be sorted and formatted in the correct style.
%
% \bibliographystyle{splncs04}
% \bibliography{mybibliography}
%

\bibliographystyle{splncs04}
\bibliography{mybibliography}
\end{document}